# 1.3kW monolithic linearly-polarized single-mode MOPA and strategies for mitigating mode instabilities

**Rumao Tao, Pengfei Ma, Xiaolin Wang[*], Pu Zhou[**], Zejin Liu**

*College of Optoelectronic Science and Engineering, National University of Defense Technology, Changsha 410073, China*
[*]*chinawxllin@163.com;* [**] *zhoupu203@163.com*

**Abstract:** We report on the high power amplification of 1064nm linearly-polarized laser in all-fiber polarization-maintained MOPA, which can operate at output power level of 1.3kW. The main amplifier was pumped with six 915nm laser diodes, and the slope efficiency is 65.3%. The beam quality ($M^2$) was measured to be <1.2 at full power operation. The polarization extinction rate of the fiber amplifier was measured to be above 94% before mode instabilities (MI) sets in, which reduced to about 90% after the onset of MI. Power scaling capability of strategies for suppressing MI is analyzed based on a novel semi-analytical model, the theoretical results of which agree with the experimental results. It shows that mitigating MI by coiling the gain fiber is an effective and practical way in standard double-cladding large mode area fiber, and, by tight coiling of the gain fiber to the radius of 5.5cm, the MI threshold can be increased to 3 times higher than that without coiling or loose coiling. Experimental study has been carried out to verify the idea, which has proved that MI was suppressed successfully in the amplifier by tight coiling.

## 1. Introduction

Many applications, such as coherent lidar system, nonlinear frequency conversion, coherent beam combining architectures, require high power linearly-polarized laser sources with single mode operation or near-diffraction-limited beam quality [1, 2]. Recently, linearly-polarized fiber laser with 1kW output power has been achieved in a monolithic fiber Bragg grating (FBG)-based Fabry-Pérot cavity [3], which employed a pairs of high power FBGs. It is a technological challenge to design FBG that can withstand multi-kilowatt power, and further power scaling may confront with some technological difficulties. Fiber laser systems based on MOPAs are typically capable of reaching high output powers, while also offering more flexibility in terms of linewidth and polarization control than a simple grating based laser [4]. Most of the high power fiber laser systems with random polarized output are based on MOPA at the moment, which has achieved output power as high as tens of kilowatt. However, power scaling of linearly-polarized MOPAs to multi-kilowatt level is currently limited by the onset of mode instabilities (MI) [4-6]. Although lot of work has been carried out to deal with MI experimentally and theoretically [7-19], few methods to mitigate MI effectively in all-fiber MOPA configuration with standard step-index large mode area (LMA) fiber have been proposed, and MI-free power scaling in all-fiber MOPA, which are based on standard step-index PM LMA fibers, is even more challenging.

In this manuscript, we present a 1.3kW-level all-fiber Yb-doped PM fiber amplifier with linearly-polarized single-mode operation. We also discuss experiments, coupled with numerical modeling, to estimate the further power scaling capability of various strategies to mitigate MI. Numerical modeling results suggest that MI-free single-mode output powers in excess of 3kW could be realized in standard step-index LMA fiber.

## 2. Experimental setup and results

A monolithic, all-fiber, Yb-doped PM fiber amplifier is shown in Fig. 1. The seed laser in the experiment is a 50 mW linearly-polarized continous-wave laser with central wavelength at ~1064nm, which was then amplified in two pre-amplifier stages to ~25W. The main amplifier consists of a 20m PM double-clad LMA Yb-doped fiber (YDF) with 21 μm diameter/0.064 NA core and 400 μm diameter/0.44 NA cladding, which is coiled loosely. Six multimode fiber-pigtailed 915 nm laser diodes were used to pump the gain fiber through a (6+1) ×1 signal/pump combiner, which can provide a maximum pump power of about 2 kW. Approximate 0.75 m long passive fiber is spliced to the gain fiber for power delivery, the output end of which is angle cleaved in order to prevent parasitic feedback from Fresnel reflection.

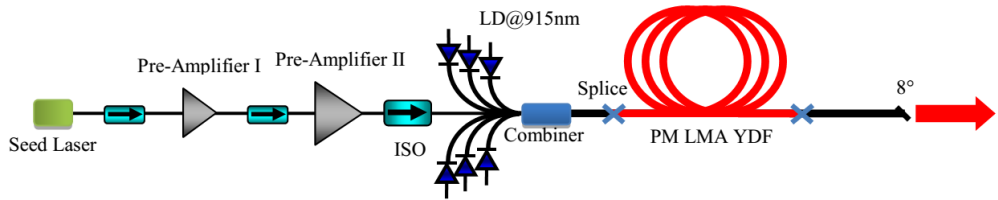

Fig. 1. The architecture of the all-fiber PM amplifier

The achieved output power at different pump power levels is shown in Fig. 2(a), which is measured after the output beam passed through a dichroic mirror. The slope efficiency of the amplifier is 65.3% with respect to launched pump power, and the pump-limited maximum output power is 1261 W. The inset picture in Fig. 2(a) shows the measured far field beam profile at the maximal operation power, and the beam quality factor $M^2$ is measured by $M^2$-200s (Spiricon) to be <1.2 in both directions, which indicates the single mode operation of the main amplifier. Fig. 2(b) shows the measured spectra (by an optical spectrum analyzer, AQ6370C Yokogawa) at full power operation, which shows that a high SNR has been achieved and no signs of parasitic lasing or significant levels of amplified spontaneous emission. Fig. 2(c) shows the measurement of polarization extinction rate (PER) for different pump power levels. It can be seen that PER is above 90% in the whole range, which indicates a linearly-polarized operation. The sudden degradation from 94% to 90% above 1.2kW was caused by MI while the slowly degradation of PER as lasing power increases may be caused by the temperature increase of the fiber.

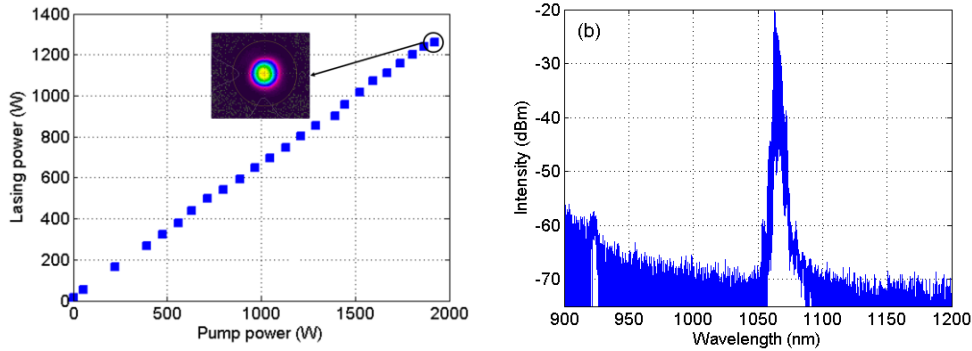

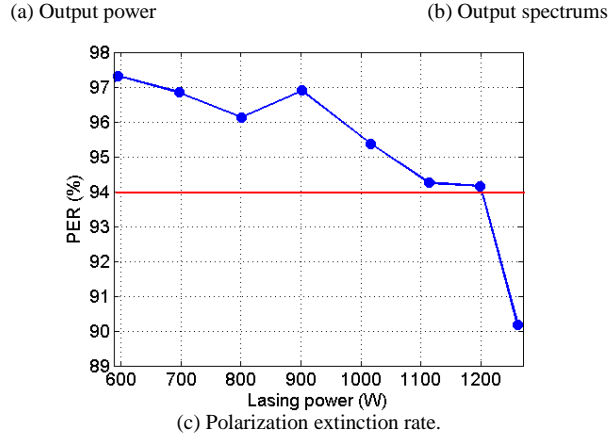

(a) Output power  (b) Output spectrums

(c) Polarization extinction rate.

Fig. 2. Output characteristics of the main amplifier.

An InGaAs photo detector (150 MHz, 700–1800 nm, Thorlabs) with a pinhole of 1.5mm diameter was put in the center of the collimated beam to monitor the onset of MI [10, 11, 19]. Time traces at different output power are shown in Fig. 3(a), and the DC component of the electrical signal is removed. From Fig. 3 (a), we can obtain that although the output power is still pump limited to 1.26kW, we are operating at the MI threshold (@1.26kW). Although the MI has set in, deterioration of beam quality has not been observed (Fig. 2(a)), which is due to that the fraction of high order mode (HOM) is relatively small at the beginning of MI [18]. It also shows that, at the onset of MI, the amplitude of the time trace is near the same as that without MI in the time period T1 and become higher than that without MI in time period T2. Applying Fourier analysis on the time traces to calculate the corresponding Fourier spectra, we obtained the frequency distribution of the beam fluctuation as shown in Fig. 3(b). The frequency components of the beam fluctuations distributed in the range of 0~200 Hz at power of 1.14kW, which means that the output beam profile is stable; however, further increasing the output power, frequency components in 0~3 kHz showed up as well as that background noise increases, which indicated the sign of instable beam profile appeared and that the amplifier are approaching the threshold MI. Similar to the observation in [20], the instability of MI has a grown-process: at the start (T1), only background noise increase, which means stable beam profile and indicate that MI may relate to noise [14]; after a few seconds (T2), frequency components shown up, which indicates unstable beam profile. After multiple power cycles near the maximum power, we observed that the threshold reduced to below 1.2kW, which may due to the fiber degradation [5, 21].

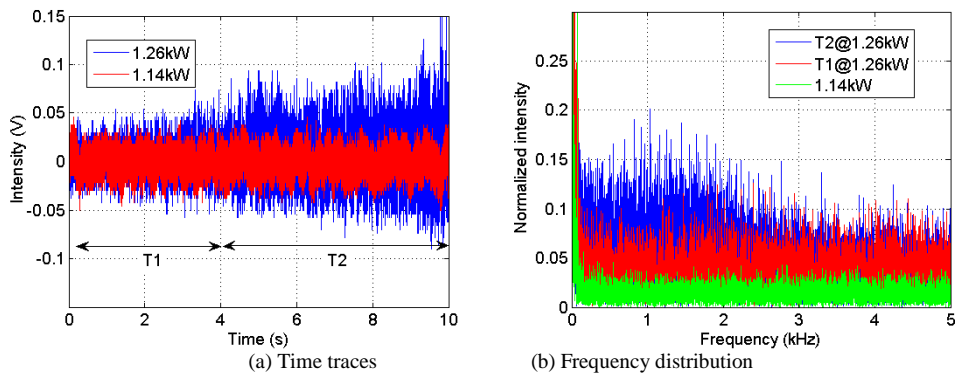

(a) Time traces  (b) Frequency distribution

Fig. 3. Fluctuation characteristics of the output beam.

## 3. Strategies for suppressing mode instabilities

*3.1 Theoretical model*

As shown in the previous section, the linearly-polarized single-mode output of fiber laser is limited by the onset ot MI. Work should be carried out to study the further power scaling of the amplifier with single-mode linearly-polarized output. One way is to carry out numerical study on MI and to find way to mitigate this phenomenon effectively. In high power fiber laser systems, most of the fibers are weakly guided fibers, where optical fields can be well approximated by linearly polarized (LP) modes. For linearly-polarized fiber lasers, the optical field of the signal propagating in the fiber is expressed in the conventional LP mode representations

$$E(r,\phi,z,t) = \sum_{m=0}^{\infty}\sum_{n=1}^{\infty} A_{mn}(z,t)\psi_{mn}(r,\phi)e^{j(\beta_{mn}z-\omega_{mn}t)} + c.c. \quad (1)$$

where $m$ and $n$ is azimuthal and radial mode numbers respectively. $A_{mn}(z,t)$, $\beta_{mn}$, and $\psi_{mn}(r,\phi)$ are slowly varying mode amplitudes, propagation constants, and normalized mode profiles of LPmn mode. Assuming the case that the fiber amplifiers are operating below or near the MI threshold, we therefore include only the fundamental mode (LP$_{01}$) and one of the two degenerate LP$_{11}$ modes and the subscripts of 01 and 11 are replaced with 1 and 2 for LP$_{01}$ mode and LP$_{11}$ mode, respectively. Then the signal intensity $I_s$ can be written as

$$I_s(r,\phi,z,t) = 2n_0\varepsilon_0 c E(r,\phi,z,t)E(r,\phi,z,t)^*$$
$$= I_0 + \tilde{I} \quad (2)$$

with

$$I_0 = I_{11}(z,t)\psi_1(r,\phi)\psi_1(r,\phi) + I_{22}(z,t)\psi_2(r,\phi)\psi_2(r,\phi) \quad (3a)$$

$$\tilde{I} = I_{12}(z,t)\psi_1(r,\phi)\psi_2(r,\phi)e^{j(qz-\Omega t)} + I_{21}(z,t)\psi_1(r,\phi)\psi_2(r,\phi)e^{-j(qz-\Omega t)} \quad (3b)$$

$$I_{kl}(z,t) = 4n_0\varepsilon_0 c A_k(z,t)A_l^*(z,t) \quad (3c)$$

$$q = \beta_1 - \beta_2, \quad \Omega = \omega_1 - \omega_2 \quad (3d)$$

The temperature distribution is governed by heat transportation equation, which is given by

$$\nabla^2 T(r,\phi,z,t) + \frac{Q(r,\phi,z,t)}{\kappa} = \frac{1}{\alpha}\frac{\partial T(r,\phi,z,t)}{\partial t} \quad (4)$$

where $\alpha = \kappa/\rho C$, $\rho$ is the density, $C$ is the specific heat capacity, and $\kappa$ is the thermal conductivity. Since the heat in high power fiber amplifiers is mainly generated from the quantum defect and absorption, the volume heat-generation density $Q$ can be approximately expressed as

$$Q(r,\phi,z,t) \cong g(r,\phi,z,t)\left(\frac{v_p - v_s}{v_s}\right)I_s(r,\phi,z,t) \quad (5a)$$

and $g(r,\phi,z,t)$ is the gain of the amplifier

$$g(r,\phi,z,t) = \left[(\sigma_s^a + \sigma_s^e)n_u(r,\phi,z,t) - \sigma_s^a\right]N_{Yb}(r,\phi) \quad (5b)$$

where $v_{p(s)}$ is the optical frequencies, $\sigma_s^a$ and $\sigma_s^e$ are the signal absorption and emission cross sections, $\sigma_p^a$ and $\sigma_p^e$ are the pump absorption and emission cross sections, $N_{Yb}(r,\phi)$ is the doping profile, the population inversion $n_u$ is given in [13].

Assume that the fiber is water cooled, the appropriate boundary condition for the heat equation at the fiber surface is

$$\kappa\frac{\partial T}{\partial r} + h_q T = 0 \quad (6)$$

where $h_q$ is the convection coefficient for the cooling fluid. By adopting the integral-transform technique to separate variables in the cylindrical system [22], Eq. (4), combined with Eqs. (5) and (6), can be solved as

$$T(r,\phi,z,t) = \frac{1}{\pi}\frac{\alpha n_2}{\eta}\sum_{v}\sum_{m=1}^{\infty}\frac{R_v(\delta_m,r)}{N(\delta_m)}$$

$$\times \int_{t'=0}^{t}\begin{bmatrix}B_{11}(\phi,z)I_{11}(z,t')+B_{22}(\phi,z)I_{22}(z,t')\\+B_{12}(\phi,z)I_{12}(z,t')e^{j(qz-\Omega t')}+B_{12}(\phi,z)I_{12}^*(z,t')e^{-j(qz-\Omega t')}\end{bmatrix}e^{-\alpha\delta_m^2(t-t')}dt' \quad (7)$$

with

$$B_{kl}(\phi,z)$$

$$=\begin{cases}\int_0^{2\pi}d\phi'\int_0^R g_0 R_v(\delta_m,r')\cos v(\phi-\phi')\frac{\psi_k(r',\phi')\psi_k(r',\phi')}{1+I_0/I_{saturation}}dr', & k=l \quad (8a)\\ \int_0^{2\pi}d\phi'\int_0^R g_0 R_v(\delta_m,r')\cos v(\phi-\phi')\frac{\psi_k(r',\phi')\psi_l(r',\phi')}{(1+I_0/I_{saturation})^2}dr', & k\neq l\end{cases}$$

$$N(\delta_m) = \int_0^R rR_v^2(\delta_m,r)dr, \quad n_2 = \frac{\eta}{\kappa}\left(\frac{v_p-v_s}{v_s}\right) \quad (8b)$$

where $v=0, 1, 2, 3…$ and replace $\pi$ by $2\pi$ for $v=0$, $\eta$ is the thermal-optic coefficient, $R$ is the radius of the inner cladding, $g_0$ is the small signal gain and $I_{saturation}$ is the saturation intensity. $R_v(\delta_m,r)$ is given by $R_v(\delta_m,r)=J_v(\delta_m r)$ and $\delta_m$ is the positive roots of $\delta_m J_v'(\delta_m R)+\frac{h_q}{\kappa}J_v(\delta_m R)=0$. Considering effective refractive index of gain from amplifier, the total refractive index, which attributes to gain ($n_g \leq n_0$) and nonlinearity ($n_{NL} \leq n_0$), can be expressed as

$$n^2 = (n_0+n_g+n_{NL})^2 \cong n_0^2 - j\frac{g(r,\phi,z,t)n_0}{k_0}+2n_0 n_{NL} \quad (9)$$

where $n_{NL}$ is given by

$$n_{NL}(r,\phi,z,t)$$
$$=\eta T(r,\phi,z,t) \quad (10)$$
$$=h_{11}(r,\phi,z,t)+h_{22}(r,\phi,z,t)+h_{12}(r,\phi,z,t)e^{jqz}+h_{21}(r,\phi,z,t)e^{-jqz}$$

with

$$h_{kl}(r,\phi,z,t) = \begin{cases}\frac{\alpha n_2}{\pi}\sum_v\sum_{m=1}^{\infty}\frac{R_v(\delta_m,r)}{N(\delta_m)}\int_0^t B_{kk}(\phi,z)I_{kk}(z,t')e^{-\alpha\delta_m^2(t-t')}dt', & k=l \\ \frac{\alpha n_2}{\pi}\sum_v\sum_{m=1}^{\infty}\frac{R_v(\delta_m,r)}{N(\delta_m)}\int_0^t B_{kl}(\phi,z)I_{kl}(z,t')e^{-\alpha\delta_m^2(t-t')-j\Omega t'}dt', & k\neq l\end{cases} \quad (11)$$

Inserting Eqs. (1) and (10) into the wave equation, after very tedious but straightforward derivations, we have obtained the coupled-mode equations

$$\frac{\partial|A_1|^2}{\partial z} = \iint g(r,\phi,z)\psi_1\psi_1 rdrd\phi|A_1|^2 \quad (12a)$$

$$\frac{\partial|A_2|^2}{\partial z} = \left[\iint g(r,\phi,z)\psi_2\psi_2 rdrd\phi+|A_1|^2\chi(\Omega,t)\right]|A_2|^2 \quad (12b)$$

with

$$\chi(\Omega) = 2\frac{n_0\omega_2^2}{c^2\beta_2}\text{Im}\left(4n_0\varepsilon_0 c\iint \bar{h}_{12}\psi_1\psi_2 rdrd\phi\right) \quad (13a)$$

$$\bar{h}_{kl}(r,\phi,z) = \frac{\alpha n_2}{\pi} \sum_{v} \sum_{m=1}^{\infty} \frac{R_v(\delta_m, r)}{N(\delta_m)} \frac{B_{kl}(\phi,z)}{\alpha \delta_m^2 - j\Omega} \quad (13b)$$

By taking the similar derivation process in [14], we can obtain the HOM content for the quantum noise (QN) induced MI from Eq. (1), which is given as

$$\xi(L) \approx \frac{\hbar \omega_0}{P_1(L)} \sqrt{\frac{2\pi}{\int_0^L P_1(z)|\chi''(\Omega_0)|dz}} \quad (14)$$

$$\times \exp\left\{\int_0^L \left[\iint g(r,\phi,z)\psi_2\psi_2 r dr d\phi\right]dz + \int_0^L P_1(z)\chi(\Omega_0)dz\right\}$$

where $L$ is the length of the gain fiber. For the other case that MI is seeded by intensity noise, we can obtain

$$\xi(L)$$
$$\approx \xi_0 \exp\left[\int_0^L dz \iint g(r,\phi,z)(\psi_2\psi_2 - \psi_1\psi_1) r dr d\phi - \alpha_{coil} L_{coil}\right]$$
$$+ \frac{\xi_0}{4} \sqrt{\frac{2\pi}{\int_0^L P_1(z)|\chi''(\Omega_0)|dz}} \exp\left[\int_0^L dz \iint g(r,\phi,z)(\psi_2\psi_2 - \psi_1\psi_1) r dr d\phi + \int_0^L P_1(z)\chi(\Omega_0)dz - \alpha_{coil} L_{coil}\right] R_N(\Omega_0)$$

(15)

where $R_N(\Omega)$ is the relative intensity noise (RIN) of the input signal, $\xi_0$ is the initial frequency shifted HOM content, $\alpha_{coil}$ is the bend-induced power loss by fiber coiling, and $L_{coil}$ is the length of the coiled length. Here bend induced loss is taken into consideration in a simple way, and the effect of bend induced mode distortion has not been considered.

*3.2 Numerical results*

In this section, we have calculated the MI threshold power of the amplifier based on the model. The parameters of the fiber is the same as those in the experiment, which are listed in Table 1.

**Table 1. Parameters of Test Amplifier**

| | |
|---|---|
| $R_{core}$ | 10.5μm |
| $R$ | 200μm |
| $n_0$ | 1.45146 |
| $NA$ | 0.064 |
| $\lambda_p$ | 915nm |
| $\lambda_s$ | 1064nm |
| $h_q$ | 5000 W/(m²K) |
| $\eta$ | $1.2 \times 10^{-5}$ K$^{-1}$ |
| $\kappa$ | 1.38 W/(Km) |
| $\rho C$ | $1.54 \times 10^6$ J/(Km³) |
| $\sigma_p^a$ | $6.04 \times 10^{-25}$ m² |
| $\sigma_p^e$ | $1.96 \times 10^{-26}$ m² |
| $\sigma_s^a$ | $6.0 \times 10^{-27}$ m² |
| $\sigma_s^e$ | $3.58 \times 10^{-25}$ m² |
| $N_{Yb}$ | $3.5 \times 10^{25}$ m$^{-3}$ |

Firstly, we calculated the MI threshold power of the amplifier. Figure 4 (a) shows the HOM content versus the pump power. It is shown that the quantum-induced-MI threshold power is about 4.7kW. However, intensity noise of the signal (RIN=$10^{-10}$ that correspond to a laser with high RIN, which yielding a realistic MI threshold [23, 24]) reduces the threshold to be about 2kW, which agrees well with the experimental results. The initial HOM is set to be

0.01. By reducing the intensity noise of the signal (RIN=$10^{-11}$), the MI threshold power can be increased about ~300W, which means that measures taken to reduce the intensity noise of the input signal result in only modest improvements in the MI threshold and adding to the overall complexity of the system [25]. The influence of HOM power was also calculated in Fig. 4 (b), which shows that the efforts to optimize the in-coupling of the signal have little impact on the MI threshold, which agrees with the experimental results [9].

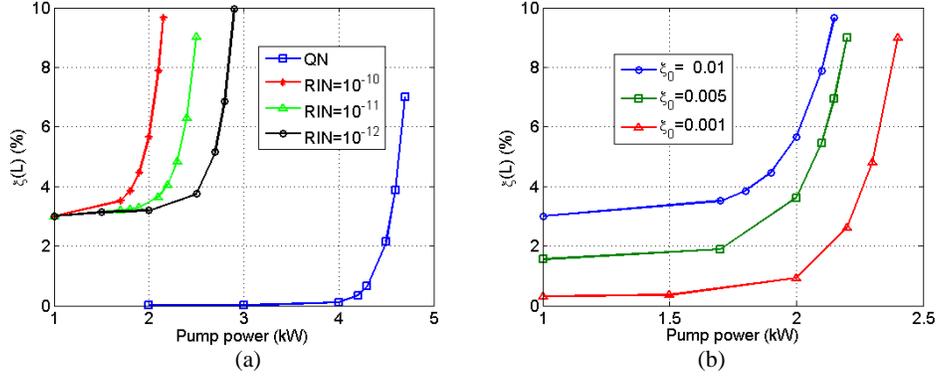

Fig. 4 Threshold calculation of the fiber amplifier.

It is reported in [26] that MI threshold power can be increased obviously by increasing cladding diameters. This is contradictive to those reported in [7], which shows that the improvement by increasing cladding diameter was not large enough for significant power scaling. To study the effect of cladding diameter on MI threshold power, we calculated the MI threshold pump power for different cladding diameters, which is shown in Fig. 5. The computed thresholds are seen to rise with increasing cladding diameters and the resulting increasing degree of population saturation [26]. However, for fiber with larger core diameters, the improvement becomes less significant, which explained the experimental results in [26]. Although MI threshold power can be improved by increasing cladding diameter, a longer length is required for fibers with larger cladding diameter to maintain good amplifier efficiency, which makes nonlinear effect suppression more challenging.

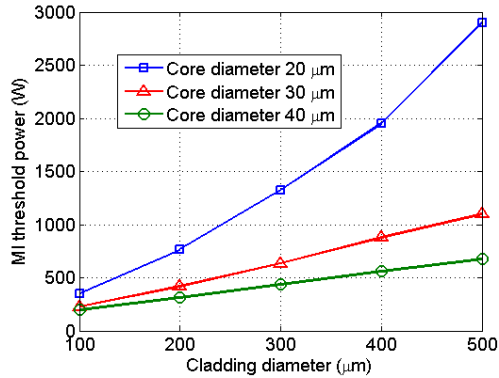

Fig. 5 Threshold pump power for different fiber cladding diameters.

Coiling the fiber with a diameter small enough to induce mode-dependent bend losses can suppress high order mode effectively [27], which can improve the MI threshold [17, 28]. The threshold power of the amplifier with tight coiling was calculated in Fig. 6. Bend losses for $LP_{11}$ mode are calculated using the method of Marcuse [29]: the bending loss is 2.4dB/m for bending radius of 6.5cm, 6dB/m for 6 cm and 14dB/m for 5.5cm. In practice, only first half of the gain fiber is coiled with small diameter, so $L_{coil}$ is set to be 6m. These coiling radiuses are chosen for long-term use. It shows that tight coiling of the fiber can increase the MI threshold power significantly: when coiling at the radius of 5.5cm, MI threshold is 3 times higher than that without coiling or loose coiling, which means that tight coiling of the gain fiber is an effective method to mitigate MI in all-fiber MOPA configuration with standard step-index LMA fiber. For the case in the experimental, MI is no longer a limitation since the onset of other nonlinear effects, such as SBS and/ or SRS, should be come into consideration first.

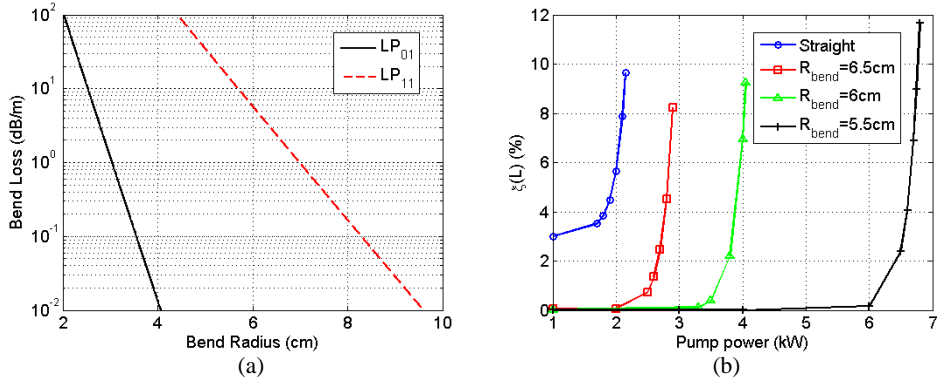

Fig. 6 The effect of coiling on MI threshold power.

*3.3 Experimental validation*

To verify our theoretical predication of the effect of coiling, we have rebuilt our fiber amplifiers with gain fiber coiled at the diameter of ~12cm. Then we have MI-free 1280W linearly-polarized single-mode laser, and all phenomenon related to the onset of MI, such as temporal fluctuation, PER degradation, have vanished (as shown in Figs. 7(a) and (b)). Due to the available pump power, the power scaling capabilities by coiling method has not been fully exploited. The advantage of employing coiling technique is that it is straightforward to implement, as most fibers are coiled in packaging anyway, and no special gain profiles are necessary to give preferential gain to fiber modes. In addition, PER of the amplifier has also been improved to ~96% at the maximal operation. The results also indicate that MI can be mitigated by designing fiber with an improved delocalization of HOM, such as chirally-coupled core (CCC) fiber [30], leakage channel fibers[31], all-solid photonic bandgap fibers [32].

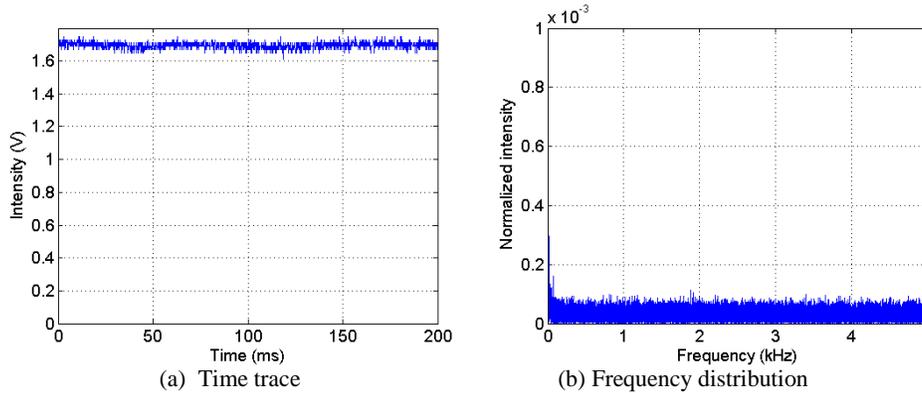

Fig. 7 Fluctuation characteristics of the output beam after tight coiling.

**4. Conclusions**

We have generated a high-power linearly-polarized single-mode output laser from an Yb-doped PM fiber amplifier, which operated at ~1064nm. The slope efficiency of the amplifier is 65.3% with respect to launched pump power, and the maximal output power is 1261W. The $M^2$ at full power operation is measured to be <1.2 in both directions. The linearly-polarization operation was deteriorated by the onset of MI above 1.2kW. The PER is measured to be >94% without MI, which reduced to about 90% after the onset of MI. A novel theoretical model to study MI has been built up, the numerical results agree well with the experimental observation. Various method to improve the power scaling capability of the amplifier without MI have been studied numerically, which reveal that MI can be suppressed by proper coiling and the amplifier in the paper have the potential to deliver MI-free 3kW output power. An additional experimental has been carried out to study the effect of coiling on MI, which rebuilt the amplifier with tighter coiling. It is shown that MI was suppressed successfully in the amplifier by tight coiling.

## Acknowledgments

The authors would like to acknowledge the support of Hanwei Zhang, Xiong Wang, Hailong Yu, Baolai Yang. The research leading to these results has received funding from the program for the National Science Foundation of China under grant No. 61322505, the program for New Century Excellent Talents in University, the Innovation Foundation for Excellent Graduates in National University of Defense Technology under grant B120704 and Hunan Provincial Innovation Foundation for Postgraduate under grant CX2012B035.